\documentclass[graybox]{svmult}
\usepackage[T1]{fontenc}
\usepackage{lmodern}
\usepackage{aas_macros} 

\usepackage{type1cm}        
\usepackage{makeidx}         
\usepackage{graphicx}        
\usepackage{multicol}        
\usepackage[bottom]{footmisc}

\usepackage{newtxtext}       %
\usepackage[varvw]{newtxmath}       

\makeindex             

\begin{document}

\title*{Terrestrial planet formation in the era of GPU computing}
\author{Simon L. Grimm\orcidID{0000-0002-0632-4407} and\\ Joachim G. Stadel\orcidID{0000-0001-7565-8622}}
\institute{Simon L. Grimm\at Department of Astrophysics, Winterthurerstrasse 190,
8057 Zurich, Switzerland, \\
Institute for Particle Physics and Astrophysics, Wolfgang-Pauli-Str. 27,
8093 Zürich,
Switzerland \email{sigrimm@ethz.ch}
\and Joachim G. Stadel \at Department of Astrophysics, Winterthurerstrasse 190,
8057 Zurich, Switzerland \email{joachimgerhard.stadel@uzh.ch}}
%
%

\maketitle

\abstract*{
 In this chapter, we summarize the underlying numerical methods needed for efficient N-body integration of planetary systems. We present state-of-the-art N-body simulations in a comparative study regarding the basic properties that emerge during the late stages of the terrestrial planet formation process. We show that in modern N-body simulations the commonly used acceleration factor f to speed up simulations should no longer be used because it can lead to different chemical composition of the planets. We compare low- to high-resolution simulations and show that the formation time scale depends on the size of the initial planetesimals. The simulations also show that terrestrial planets can form resonant chains without the need of orbital migration due to gas effects. 
}

\abstract{
 In this chapter, we summarize the underlying numerical methods needed for efficient $N$-body integration of planetary systems. We discuss how symplectic integrators have been developed to tackle the complementary problems of long-term orbital integration and short-term collisional interactions. The public code GENGA, a parallel GPU/CPU planet formation and orbital dynamics simulation code, was developed to unify these methods and take full advantage of the newest available computing hardware. We present state-of-the-art $N$-body simulations performed with GENGA in a comparative study regarding the basic properties that emerge during the late stages of the terrestrial planet formation process. We show that in modern $N$-body simulations the commonly used acceleration factor $f$, used to speed up the collisional growth of planets in simulations, should be avoided since it can lead to distorted chemical composition of the planets. We make a detailed comparison of low to high-resolution simulations, showing that the formation time scale depends on the size of the initial planetesimals. These simulations also show that terrestrial planets can form resonant chains without the need of orbital migration due to gas effects. 
}

\section{Introduction: Planet formation}

There are several stages in the process of planet formation \cite{Armitage2007} and the physics of every stage is a complex topic itself, and each stage is dominated by different physical effects and time scales. Starting with the formation of a protoplanetary disk containing gas and dust particles around a young star. The dust particles in the disk begin to stick together and form larger objects, grains, clumps, and finally planetesimals \cite{Raymond+2022, Raymond+2020, Drazkowska2023, Clement+2024, Armitage2024}. From there, a stage of protoplanet formation follows, where planetesimals collide to form larger bodies \cite{Raymond+2013, Clement+2024}. The bodies are still embedded in the gas disk, which leads to orbital migration effects. 
The last stage is the long-term evolution phase, at which point the gas disk has mainly dissipated and is no longer significant to the dynamics of the bodies. Collisions between objects, leading to planetesimal accretion or even giant impacts. Applied to the Solar System, there are several common theories/scenarios for the formation of planets. For example, the Nice model \cite{Tsiganis2005}, the Grand Tack model \cite{Walsh+2011}, planet formation from a small ring of planetesimals \cite{Drazkowska+2016, Izidoro+2022, Woo+2023}, or planet formation from pebble accretion \cite{Johansen"2021}. To what extent these models apply to extra-solar planetary systems remains an open question.

The process of terrestrial planet formation strongly depends on the architecture of the planetary system, mainly on whether there are gas giants present or not. When gas giants are present, they can deliver material to the inner part of the system and accelerate the formation process of inner terrestrial planets. When no gas giants are present, the formation process can take much longer, but in contrast to the Solar System, these planets are much more similar in size and regularly spaced in the orbital separation \cite{Weiss+2018} or even appearing in mean motion resonance chains. On the other hand, when the included gas giants are too massive, they can completely suppress the formation of terrestrial planets by scattering all available material out of the system. 

In this chapter, we cannot address all aspects of planet formation, but instead we focus on those where present $N$-body simulation methods can change or improve our understanding, namely the processes dominated by gravitational interactions. This is the middle- to late-stage evolution of the planetary system, starting with a distribution of planetesimals, and ending billions of years later with the final configuration of the formed planets.

We start with already formed planetesimals and study the evolution up to the final planets.

\subsection{From planetesimals to planets}

Planetesimals are considered to be the building blocks of terrestrial planets. They are exposed to mutual gravitational forces from all other bodies in the system, leading to $N^2$ gravitational force terms. While the bodies attract each other, collisions between them form more massive bodies, which in turn attract even more bodies. This process is expressed in the theory of runaway growth, which describes an accelerated phase of planetary growth, followed by slower oligarchic growth \cite{KokuboIda1998, Chambers2006}. During close encounters, larger bodies can transfer kinetic energy and angular momentum to surrounding smaller bodies, a process called dynamical friction or gravitational drag. This process dampens the eccentricities of the larger bodies, causing orbital migration, while the small bodies are set to more eccentric orbits. A consequence of dynamical friction is that the growing and migrating planets can be trapped in mean-motion resonances, thereby forming resonant chains of multiple planets.

The entire gravitational interaction between all bodies cannot be easily described analytically and needs to be studied with $N$-body simulations. 

A classic $N$-body simulation for terrestrial planet formation starts with a disk of $N_{\rm initial}$ planetesimals, distributed in a disk around a central star. The planetesimals can either have a size distribution or all have the same size and mass. The total mass as well as the radial mass distribution generally follow the Minim Mass Solar Nebula (MMSN) estimation \cite{Hayashi1981, KokuboIda2002, Morishima2010}. Initially, the planetesimals have nearly circular and coplanar orbits around the central star. Close encounters between bodies lead to excitement in the eccentricity and inclination, giving a random velocity distribution relative to a perfect Kepler orbit. The process of viscous stirring further increases the random velocities.
Thus, the system can be described as a collisional stellar-dynamical system. Similarly as in two-body relaxation in star clusters, the bodies redistribute energy and angular momentum to the other bodies.

As time evolves, planetesimals begin to collide and form larger bodies. Since the orbital velocity and typically also the particle density of the bodies are larger in the inner part of the system than in the outer part, bodies grow first in the inner part. In the outer part, it takes much longer, but finally, the bodies will also form larger objects there. Very similarly to the process of terrestrial planet formation through $N$-body dynamics is the formation of cores of gas giants. But a major difference is that in core formation simulations a detailed and realistic gas disk description is much more important, making it numerically even more challenging for long-term simulations. Therefore, we restrict ourselves to the case of the formation of terrestrial planets.

Pioneering work on such $N$-body simulations has been done in \cite{KokuboIda1998, KokuboIda2002}. Since these $N$-body simulations are expensive to run, often an artificial acceleration of the growth process is implemented by increasing the physical radii of the bodies by a factor $f$ \cite{KokuboIda1996}. With this factor, the growth time scale can be reduced approximately by $f^{-2}$, depending on the growth mode \cite{KokuboIda1996}. Today, the available computational hardware is powerful enough to run simulations without this speed-up factor.
In Section~\ref{sec:f-factor}, we present a new analysis on the convergence and speed-up resulting from the use of this approach and demonstrate that it should be avoided when correctness in the assembly, particularly the feeding zone, of the terrestrial planet is of importance.

\section{The $N$-body problem}

In planetary science, there are two main uses of $N$-body simulations: precise short-term predictions of celestial bodies, such as forecasting the position of an asteroid in the Solar System over the next few decades \cite{Marsden1973, Sitarski1983}, and long-term simulations of an $N$-body system spanning millions or billions of years. In the first case, the positions of the objects of interest must be calculated as precisely as possible over a relatively short time span. Adaptive time-step methods such as the high-order Runge-Kutta-Fehlberg \cite{Hairer1993} or similar methods are well suited for this task. In the latter case, maintaining conservation laws, such as energy conservation, is more important than pinpointing the exact locations of individual objects, and different numerical methods must be used. In this chapter, we will focus only on the second part and apply it to the problem of terrestrial planet formation via planetesimal accretion.

As available hardware improves over time, new parallel computational techniques are also developed, and we must constantly refine our methods to make the best use of the available devices. Over the last 10 years, especially Graphics Processing Units (GPUs) have shown an enormous boost in processing power. We will show how modern hardware, combined with state-of-the-art numerical methods, improve the quality of $N$-body simulations, and we will also show why the development must still be pushed forward for even more detailed simulations to reveal new physical insights.

\subsection{Solving the $N$-body problem}

The $N$-body problem in celestial mechanics must solve the dynamics of $N$ bodies under the influence of the Newtonian gravitational force. The very fascinating fact is that this underlying physical formula is very simple and consists of only a handful of mathematical operations, and a basic numerical method can be implemented in a relatively small code. By doing that, it would become apparent very soon that such a naive implementation would suffer from performance and accuracy problems, showing that advanced and optimized numerical methods are needed for just a simple equation.

Since there are in total $N$ bodies in the system, the $N$-body problem consists of a double sum in all terms of mutual forces, and hence a computational complexity of $\mathcal{O}(N^2)$.
As we are mostly interested in a regime of many thousands of bodies, the computational costs can easily explode and exceed a reasonable computational time. This is the main reason why the $N$-body problem is so hard to solve for large values of $N$. Another challenge is the wide range of dynamical time scales. Typical time steps can vary from minutes during a close encounter to years for distant planetary Kepler orbits.

\subsection{The symplectic integrator}

We are interested in a numerical method that can simulate the process of terrestrial planet formation over millions or billions of years. For this application, it is essential that the energy of the system is conserved well. Otherwise, the planets would start to spiral in- or outward and leave the system at some point in time, or even cause unnatural collisions. That is a clearly unphysical behavior and must be avoided. A numerical integrator that fulfills energy conservation is the symplectic integrator. This method splits the integration into two independent parts by separating the physical variables. The integrator then solves each part independently and updates them in an interleaved way. A well-known example of such an approach is the 'leap-frog' integrator, which separates the system into kinetic and potential terms. Written in terms of a Hamiltonian, it takes the form:

\begin{equation}
    \label{eq_leapfrog}
    H = H_{\rm kinetic} + H_{\rm potential}.
\end{equation}

This integration scheme has been used very successfully in many general $N$-body simulations. However, for planetary systems, we can optimize the integrator and take advantage of the fact that the most dominant contribution to Equation \ref{eq_leapfrog} is the gravitation of the central mass. The gravitation from all other bodies is typically much smaller, unless they are in a close approach. That means that most bodies orbit the central star on Keplerian orbits, which are slightly perturbed by the presence of additional bodies. Since a Keplerian orbit can be solved analytically, this allows us to use much larger time steps than in a leap-frog integrator, and therefore speed up the integration time significantly. In \cite{WisdomHolman1991} such an integrator is described by formally splitting the Hamiltonian of the system into a Keplerian part and an interaction part: 

\begin{equation}
    H = H_{\rm kepler} + H_{\rm interation}
\end{equation}

A separation that requires somewhat complicated Jacobi coordinates and is limited to the case where perturbations from other bodies remain much smaller than the force from the central star.

\subsection{The hybrid symplectic integrator}

A major disadvantage of the symplectic integrator in \cite{WisdomHolman1991} is that it cannot handle close encounters between bodies accurately enough. An essential aspect in planet formation simulations, where close encounters and collisions can occur frequently.

During a close encounter, the mutual gravitational force between the involved bodies can exceed the gravitational force of the central star. Therefore, the perturbation approach by splitting the system in Keplerian and interaction parts breaks down, and the integrator would start to make large errors in the orbits. Therefore, a better integration scheme is needed.

A solution to the problem was found by \cite{Chambers99} with the introduction of the hybrid symplectic approach and by using mixed coordinates, also called democratic coordinates \cite{Duncan1998}. These are heliocentric positions $\vec{r}_i$ and barycentric velocities $\vec{v}_i$. The idea of this integration method is to isolate the close encounter phases from the symplectic integration method and to integrate them with a different integrator, which is able to refine the time step. The separation of these terms must be done smoothly enough to keep the integration error to a minimum. The Hamiltonian of the system then takes the form:

\begin{equation}
    \label{eq_H}
 H = H_{A} + H_{B} + H_{C},
\end{equation}

with 
\begin{eqnarray}
\label{Ha}
H_{A} = \sum_{i=1}^{N-1} \left( \frac{p_{i}^{2}}{2m_{i}}  - \frac{G m_{i} m_0}{r_i} \right) \nonumber \\
- \sum_{i = 1}^{N-1} \sum_{j = i+1}^{N-1} \frac{G m_{i} m_{j}}{r_{ij}} [ 1 - K(r_{ij})],
\end{eqnarray}

\begin{equation}
\label{Hb}
H_{B} = -\sum_{i = 1}^{N-1}\sum_{j=i +1}^{N-1} \frac{G m_{i} m_{j} }{r_{ij}} K(r_{ij}),
\end{equation}
and
\begin{equation}
\label{Hc}
 H_{C} = \frac{1}{2m_0}\left( \sum _{i =1} ^{N} \texttt{p}_{i} \right) ^{2}.
\end{equation}

$K(r_{ij})$ is a smooth changeover function ranging from 0 to 1, which separates close encounter phases from the rest of the system. A critical radius describes the separation distance between the involved bodies, when the changeover starts to act. This radius depends on the Hill radii $R_{\rm Hill}$ of the bodies involved, scaled by a numerical safety factor, usually a factor of 3. Since the transition from the symplectic integrator to the direct integrator must be done smoothly, that is, over some minimal number of time steps, a second term is used to increase the critical radius in situations where too few steps are used through the changeover function. This second term includes the distance a particle can usually travel within a time step towards another particle and is given as the time step $dt$ times the velocity $v$ of the particles times another numerical safety factor - usually 0.4. In this way, the critical radius is the maximum of the two described terms and is given as \cite{Chambers99}:

\begin{equation}
    \label{rcrit}
    r_{\text{crit},i}= \max(3 \cdot R_{\rm Hill,i}, 0.4 \cdot dt \cdot v_i).
\end{equation}

The part $H_A$ describes Keplerian orbits. When no close encounters occur, this part can be solved analytically; when there is a close encounter, a direct solver is needed which propagates the involved bodies through the close encounter phase. The part $H_B$ describes the $N^2$ mutual gravitational force terms. The part $H_C$ arises as a consequence of using mixed variables and describes the total momentum of the system. The inner direct solver in the close encounter phases must not necessarily be symplectic, because these phases are relatively short. However, it is important that the transition between the outer symplectic integrator and the inner direct integrator happens smoothly enough.

This hybrid symplectic integration method is implemented in the Mercury code \cite{Chambers2012} and is very successful in simulating the formation of terrestrial planets, and has been used in a large number of scientific publications over many years. Other commonly used codes are SyMBA \cite{Duncan+1998}, REBOUND \cite{Rein+2012}, or GPLUM \cite{Ishigaki+2021}.

\subsection{Increasing $N$}
\label{sec:N}

The hybrid symplectic integrator described above and implemented in the Mercury code \cite{Chambers2012} works very well for system sizes with $N <\approx 2000$. This limitation in the number of particles also limits the physical completeness of the simulations. In this regime of $N$, the initial body sizes are not much smaller than the Earth's moon. Certainly not a realistic planetesimal size if one does not start already with some planetary embryos in the system. Another limitation is that the particle resolution is not sufficient to analyze detailed chemical transport and composition within the system or to include an asteroid belt. In Section \ref{sec:compareN} we analyze the difference between low and large $N$ simulations in more detail.

When a larger number of particles is used in a planetary system simulation, it can happen that the bodies get too tightly packed, such that every particle is in a close encounter with another particle. This would lead to a configuration where the outer symplectic integrator has no impact, and all work is done with the inner direct integrator that resolves the single group of close encounters that contain all particles \cite{GrimmStadel2014}. The only solution to this would be to reduce the time step of the symplectic integrator, which would slow down the integration. The situation can be improved by finding independent close encounter groups and integrating them independently, or by introducing higher-level changeover functions \cite{Grimm+2022}. The idea is to not switch from the symplectic integrator directly to the direct integrator but to introduce intermediate phases, where the symplectic integrator reduces the time step level by level for the affected bodies. Only the last level is then performed with the inner direct integrator. Every level needs again a smooth changeover function to reduce the integration errors of the time-step change. This method is implemented in the GENGA code \cite{GrimmStadel2014,Grimm+2022}. A similar idea is implemented in SyMBA \cite{Duncan1998}. Although GENGA is able to integrate systems with up to $\approx$ 60'000 fully interactive particles in a reasonable amount of time with this method, there is interest in integrating even larger systems. 

The range of $N$ around 60'000 is computationally very interesting, because it is roughly the range where the direct $N^2$ summation of all forces can be replaced with a more complex force solver like the Fast Multipole Method (FMM) \cite{Greengard_Rokhlin_1997, Dehnen2002,Yokota2010}. The idea of FMM is to organize the system in a grid of cells, which can be either adaptive or static, and to approximate the gravitational acceleration of each cell by a multipole expansion. The multipole expansion terms of different cells can be combined in an efficient way, leading to a complexity of just $\mathcal{O}(N)$, but the method comes with a more complicated data structure. And the high algebraic complexity due to the high order of the needed expansions adds overhead time to the calculations.

\subsection{Modelling additional dynamical effects}

There are some additional physical forces and effects that are important for the formation of terrestrial planets, in addition to pure Newtonian gravity. In the following section, we list the most important ones. Most of these effects depend on both the position and the velocity of the bodies and cannot be directly added to the symplectic integrator since it would violate energy conservation. However, it is possible to add acceleration due to these additional forces, $\vec{a}_{f}$, with an implicit midpoint method \cite{Hairer1993}: 

\begin{equation}
\label{eqn:implicit}
\vec{v}^{t + dt}_i = \vec{v}^{t}_i + dt \cdot \vec{a}_{f}\left(\vec{r}^t_i, \frac{\vec{v}^{t}_i + \vec{v}^{t + dt}_i}{2} \right),
\end{equation}
with the time step $dt$ and the time index $t$. The implicit midpoint method needs some iterations to converge and is applied in between of the basic symplecitc operations \cite{GrimmStadel2014}. Detailed equations for the following effects can also be found in \cite{GrimmStadel2014}.

\bigskip
\textbf{Gas effects:}
Typically, planet formation simulations start at a time when a gas disk is still present in the system. This disk can have a very important impact on the dynamics of the bodies.
The effects of the gas disk include gas drag, an additional gravitational force from the mass of the disk, and tidal interaction between the planet and the gas disk \cite{Morishima2010}. The implementation of the effects of the gas disk is not trivial and ranges from simple one-dimensional analytical models as in \cite{Morishima2010} to full three-dimensional hydrodynamic simulations as in \cite{Benitez2016}. However, since the physical time scales of the gas effects are much shorter than the time scales of the $N$-body problem, often the gas model can be kept relatively simple.
As the gas disk dissipates over time, its influence on the dynamics typically vanishes after a couple of million years.

\bigskip
\textbf{General relativity:}
Near the central star, general relativity is causing orbital precession of the planets. Typically, full general relativity is not implemented in $N$-body codes. Post-Newtonian corrections \cite{Kidder1995, Fabrycky+2010} provide adequate accuracy for planet formation and asteroid orbit calculations.



\bigskip
\textbf{Tidal forces and rotational deformation:}
When bodies are not considered as point masses, tidal forces are important \cite{Hut1981}. Mostly tidal forces between the star and the planets are considered, but if necessary tidal forces between bodies can also play a role. When the star rotates and is described by an oblate ellipsoid instead of a perfect sphere, rotational deformation forces must be taken into account \cite{Moyer1971}. Both of these forces add a torque on the star and the rotating bodies, changing their spin.

\bigskip
\textbf{Yarkovsky effect:}
 The Yarkovsky effect is important for the orbits of small bodies such as asteroids or fragment particles. This effect acts on rotating bodies and is a consequence of the fact that bodies are radiated on one side from the central star and emit radiation in another direction due to rotation \cite{Bottke+2006, Broz2006}. The Yarkovsky effect is very small, but sufficient to drift small bodies slowly into mean motion resonances of larger planets and, therefore, cause particles to be scattered into different parts of the planetary system. A similar effect to the Yarkovksy effect is the YORP effect \cite{Rubincam2000}, which is a consequence of irregular shapes or irregular surface properties of the bodies and can also modify their spin rates.

\bigskip
\textbf{Poynting-Robertson effect:}
When even smaller particles, such as dust particles, are considered, the Poynting-Roberston effect causes a radial drift of the particles \cite{Burns1979}. Similarly to the Yarkovsky effect, this force is very small but can cause dust pile-up effects in resonance traps \cite{Kortenkamp+2013}.

\bigskip
\textbf{Detailing the outcomes of collisions:}
$N$-body simulations often treat collisions as perfect mergers. This is not what happens in reality, where collisions can lead to the creation of fragment particles. Although individual collisions between bodies can be simulated with SPH codes, for example with PKDGRAV3 \cite{Potter+2017, Meier+2025}, it is not practical to do this at every collision in an $N$-body planet formation simulation. However, it is possible to use SPH collision simulations to create a database, from which the $N$-body code can generate fragment particles on its own \cite{Genda2017, Timpe+2020}.

\bigskip
All of these effects are implemented within GENGA II \cite{Grimm+2022}.

\subsection{The hardware situation}
\label{subsec:Hardware}

In $\sim$ 2014 (corresponding to the start of the NCCR PlanetS) an exciting era in computational hardware was beginning. It became possible to run scientific software codes using Graphics Processing Units (GPUs) by using NVIDIA's programming language CUDA. This new opportunity to potentially speed up scientific simulations has led to a GPU hype. Supercomputer centers have begun to install new hybrid systems with GPUs. The Swiss National Supercomputing Centre (CSCS), for instance, has upgraded in 2013 the machine \verb|Piz Daint| to the first hybrid architecture by adding Nvidia GPUs to the CPU nodes.
Next to the big supercomputing centers, it was also attractive for smaller institutions to invest in GPU hardware. At that time, it was especially interesting to install the less expensive, consumer grade, gaming cards. These gaming cards initially offered the same computing power as the professional data center GPUs, both in single precision (32-bit) calculations as well as in double precision (64-bit) calculations. Later consumer grade models no longer provided this level of performance for double precision calculations (which are most relevant for general scientific software), but retained the excellent single precision performance. 

Although the hardware situation at that time looked very promising, there was still a lack of scientific software that could actually use the compute capabilities efficiently. It turned out that the usage of GPUs was not as straightforward as assumed. GPU codes need to closely respect the hardware architecture. In particular, the correct usage of the different memory types can be challenging. Often, GPU codes need to be reimplemented from scratch and cannot simply be ported from a CPU code. The development of GENGA provided the community with the first efficient GPU $N$-body code for planet formation simulations.

Within the last ten years, the computing power of GPUs has continued to increase dramatically, and almost all supercomputers now make use of them. The reason for this is certainly the rise of machine learning algorithms and artificial intelligence, which is currently revolutionizing the field of computing and driving the GPU development even further.
Huang's law states that the processing power for int8 operations in GPUs should double every two years. In simulations using FP32 or FP64 operations instead of int8, this growth in speed is a bit less, but still very impressive. Basically, it means that a simulation that originally took half a year to complete can now be run within a week.

In addition to the very impressive development of GPUs, one should not forget the progress made on the CPU side. Especially AMD multicore processors with clock rates up to 5 GHz can outperform GPUs in regimes where the simulation cannot make use of the full GPU. This is the case in small to intermediate-sized $N$-body simulations, as illustrated in Figure \ref{fig:Performance}. Again, another improvement that happened in the hardware was the incorporation of advanced vector extensions AVX-256 and AVX-512 in CPUs which allows parallel execution of floating point calculations, narrowing the gap between CPUs and GPUs somewhat for floating point calculations. Interesting is also that developments for GPU parallelization can be port more easily to vector extension on the CPU.

\subsection{The GENGA code}

A major limitation of the Mercury code is that it is not parallelized. Since the calculation of the mutual gravitational force between all particles increases with $\mathcal{O}(N^2)$, the simulation run time with more than a few hundred bodies can easily take too long to be feasible. Another problem is that by using more and more particles, close encounter pairs start to cluster together and form larger and larger close encounter groups as described in Section \ref{sec:N}. 
In order to be able to run larger simulations and speed up the calculations using GPUs, the GENGA code was developed \cite{GrimmStadel2014,Grimm+2022} \footnote{https://bitbucket.org/sigrimm/genga}. 

Although GPUs are extremely powerful for large $N$ simulations, they suffer from poor performance when not enough particles are used to fill the entire device. For this small $N$ regime, a parallel CPU integrator is fastest. Furthermore, $N$-body simulations can start with many bodies, but lose them constantly over time due to collisions or ejections. Therefore, the code needs to use both parallelization techniques, GPU for large $N$ and parallel CPU calculations for small $N$. 
The newest version of GENGA supports both of these parallelization strategies, automatically translating the GPU programming language CUDA to C++ and OpenMP for the small $N$ regime. GENGA also provides a translator from CUDA to HIP, the programming language of AMD GPUs.

\begin{figure}[ht]
\includegraphics[width=\textwidth]{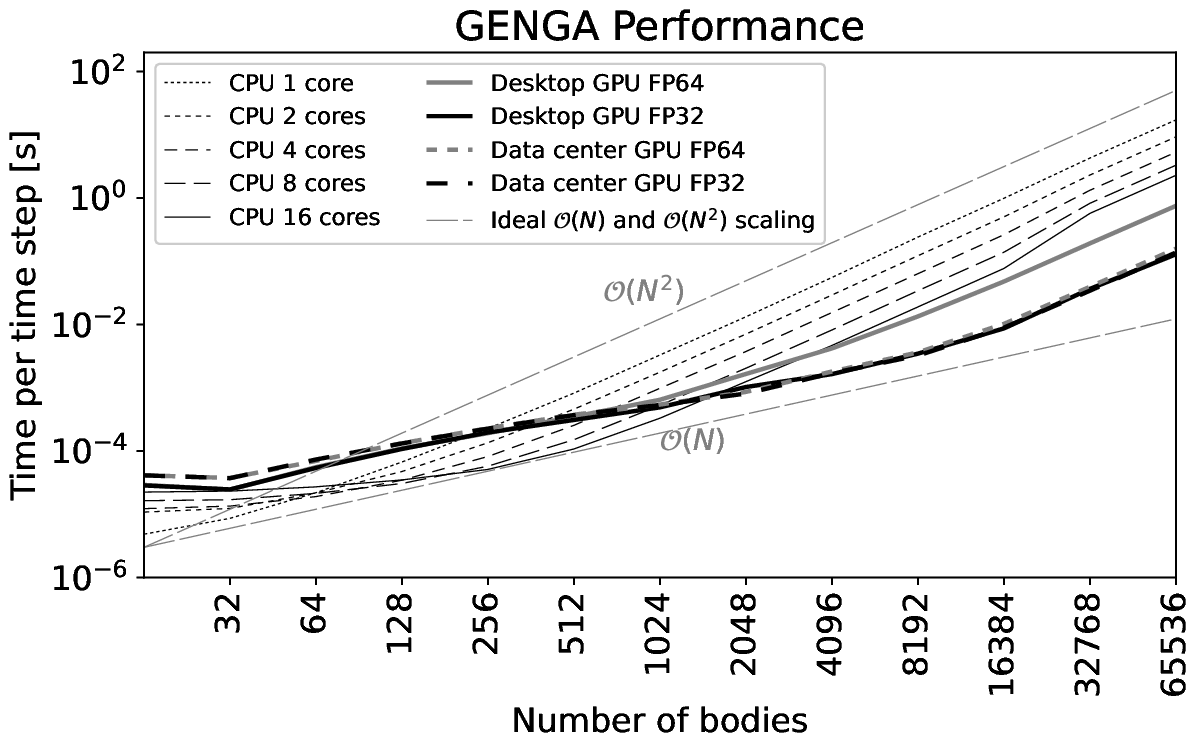}
\caption{Performance of the GENGA integrator for a disk of $N$ planetesimals. For large $N$, the GPU is clearly faster than CPU calculations. For small $N$, a multicore CPU can beat the GPU. Clearly visible is the difference in double precision (FP64) versus single precision (FP32) performance of the desktop GPU. A data center GPU does not have such a large difference. Used are a Nvidia Geforce RTX 4070ti GPU with 7680 CUDA cores and a clock speed of 2.3 GHz, a Nvidia data-center Tesla P100 with 3584 CUDA cores and a clock speed of 1.3 GHz, and an AMD Ryzen 9 7950X3D 16-Core Processor CPU with a clock speed of 4.2 GHz. In grey dashed lines are shown the idealized linear and quadratic performance scaling.}
\label{fig:Performance}
\end{figure}

In Figure \ref{fig:Performance} the performance of GENGA is shown for a simulation containing $N$ bodies distributed in a disk between 0.5 AU and 4 AU and containing a mass of $5 M_{\oplus}$ orbiting the Sun according to the minimum mass Solar nebula (MMSN). These results show that a GPU can speed up large simulations by a factor of 150 with respect to a single core CPU. Compared to using 16 CPU cores, it is still a factor of 20 faster. For small simulations, it is very interesting to see that a multicore CPU can beat the GPU at least by a factor of 10. This is due to the slower clock speed of the GPU, the larger kernel overhead time, and a lower memory transfer rate to the GPU. The figure shows very clearly how important it is for an $N$-body integrator to be able to use different hardware and parallelization techniques for different problem sizes. For this reason, GENGA makes use of some self-tuning routines to figure out the fastest integration setup.

\section{New results from GPU $N$-body simulations}

After an introduction to the $N$-body method and hardware development, we present new results in three comparative studies. These studies were only possible to run with the superior performance of GPUs. First, we test the influence of the acceleration factor $f$ used in \cite{KokuboIda1996},  then we compare low to high resolution simulations for simulations without and with gas giants. All simulations presented are run with GENGA II \cite{GrimmStadel2014}.

\subsection{Testing the acceleration factor $f$}
\label{sec:f-factor}

The acceleration factor which artificially increases the radius of the bodies by a constant factor, $f$, was originally introduced to speed-up the collision frequency in the simulation, and therefore accelerate the entire physical time scale by $\approx f^2$ depending of the growth mode (run-away or oligarchic) \cite{KokuboIda1996}. Today, the available computing hardware is powerful enough to run simulations without this factor and it is rarely used in current publications. However, some studies still use this method \cite{Wallace2019, Nader+2025}. It is not clear what physical consequences result from its use in long-term simulations. In this section, we want to demonstrate the impact of the acceleration factor in N-body simulations.

\begin{figure}[ht]
\includegraphics[width=\textwidth]{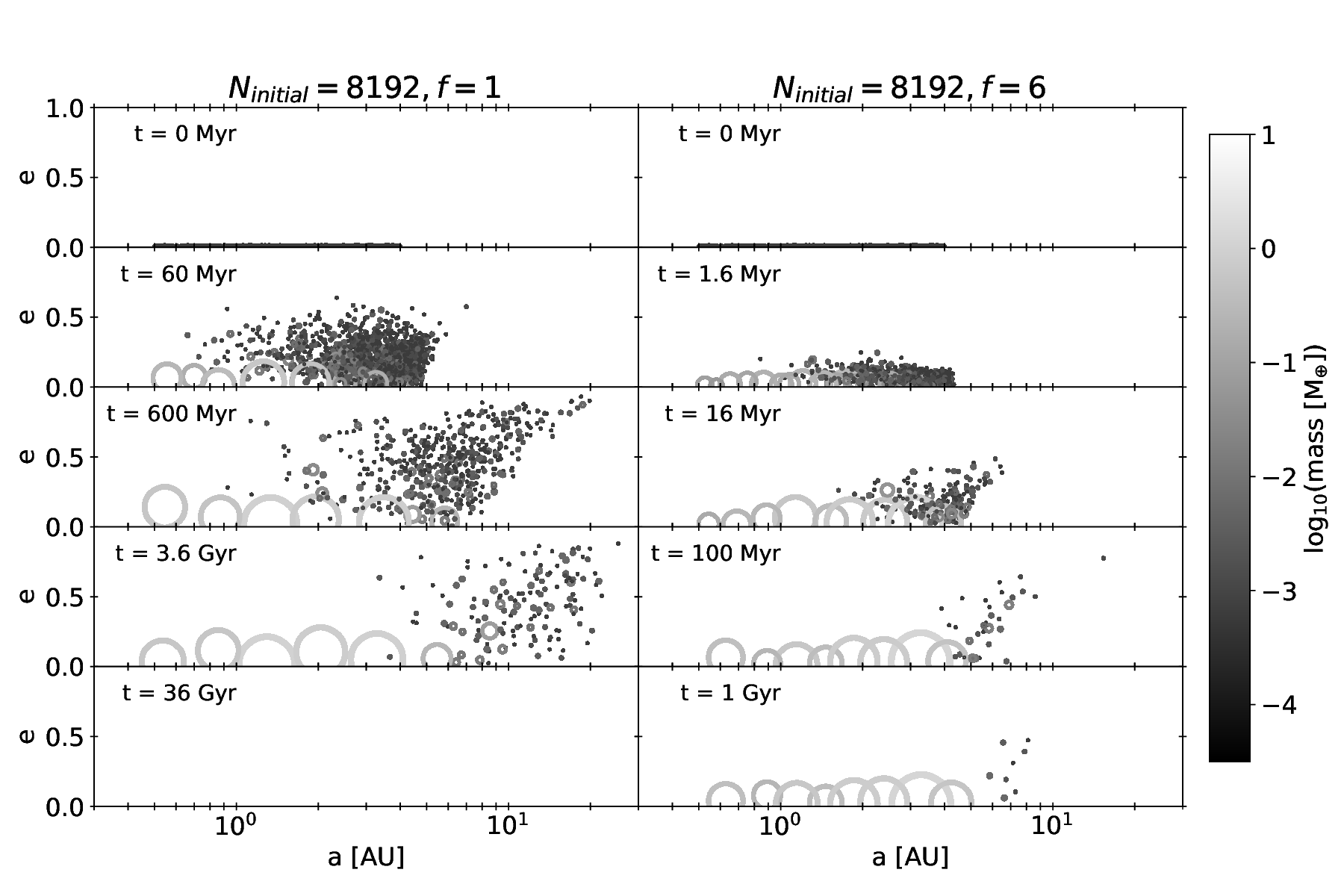}

\caption{Testing the acceleration factor $f$ for $N$-body simulations of the formation of terrestrial planets. Both simulations start with 8192 planetesimals, distributed in a disk between 0.5 and 4 AU and a total mass of $5 M_{\oplus}$. Shown are the eccentricity $e$ versus the semi-major axis $a$ for different snapshots in time $t$. The color and the size of the symbols represent the mass of the bodies. The acceleration $f$ reduces the growth time roughly by $f^{-2}$, but the final outcome show differences in the spacing between the formed planets. Also small bodies appear to also be dynamically hotter. The bottom left panel is empty, because the integration was performed only over 4 Gyr.}
\label{fig:ae_f6}
\end{figure}

To test the impact of the acceleration factor $f$, we performed $N$-body simulations with $f = 1$ and $f = 6$, and analyze the differences. The results of two such simulations are shown in Figure \ref{fig:ae_f6}. We start with $N$ equal-sized planetesimals, distributed between 0.5 and 4 AU, with a total mass of five Earth masses and no included gas disk. Due to the high orbital velocity at the system's inner edge, frequent close encounters between bodies occur, leading to the initial formation of planets at the inner boundary and subsequently towards the outer regions.

A first notable effect of the acceleration factor is that at the corresponding snapshots in time, the eccentricities of the small bodies of the accelerated simulations remain much smaller, and the systems remain also more compact in the semi-major axis. 

The second effect on an increased $f$ value is in fact a speed-up in the growth time scale of the formed planets by approximately a factor of $f^{-2}$. The simulation state of 3.6 Gyr can be achieved already after 100 Myr. At these snapshots in time, both simulations have formed a resonant chain of nearly equal-sized planets between 0.5 and 5 AU, but there are differences in the number of formed planets and the spacing between them. This difference is a direct consequence of the lower eccentricity values in the accelerated simulation, which leads to fewer orbital crossing configurations with the growing planets and a reduced feeding zone of the planets. 
In Table \ref{tab:P} are listed the period ratios between all subsequent planet pairs in the final time snapshot. While the simulation with $f = 1$ shows many pairs of planets in a strong 2:1 resonance configuration, the simulation with $f = 6$ consists of more compact and weaker resonances. It is important to note that this described effect is visible not only in the two particular simulations shown here but also in more instances of similar simulations and is not just a consequence of the chaotic nature of the $N$-body problem.

\begin{table}[ht]
\caption{Orbital period ratios of the final planets shown in Figure \ref{fig:ae_f6}. All formed planets are in orbital mean motion resonance configurations. The simulation with $f=1$ has typically stronger resonance configurations. Similar results are seen in additional simulations not presented in this work, but to prove a generic global result more simulations would be necessary.}
\label{tab:P}

\begin{tabular}{p{2cm}p{1cm}p{1cm}p{1cm}p{1cm}p{1cm}p{1cm}p{1cm}}
\hline\noalign{\smallskip}
Planet pairs & 1-2 & 2-3 & 3-4 & 4-5 & 5-6 & 6-7 & 7-8\\
\noalign{\smallskip}\svhline\noalign{\smallskip}
$f = 1$ & 2:1 & 11:6 & 2:1 & 2:1 & 2:1 & - & -\\
$f = 6$ & 5:3 & 3:2 & 3:2 & 3:2 & 3:2 & 8:5 & 3:2\\
\noalign{\smallskip}\hline\noalign{\smallskip}
\end{tabular}
\end{table}

\begin{figure}[ht]
\includegraphics[width=0.5\textwidth]{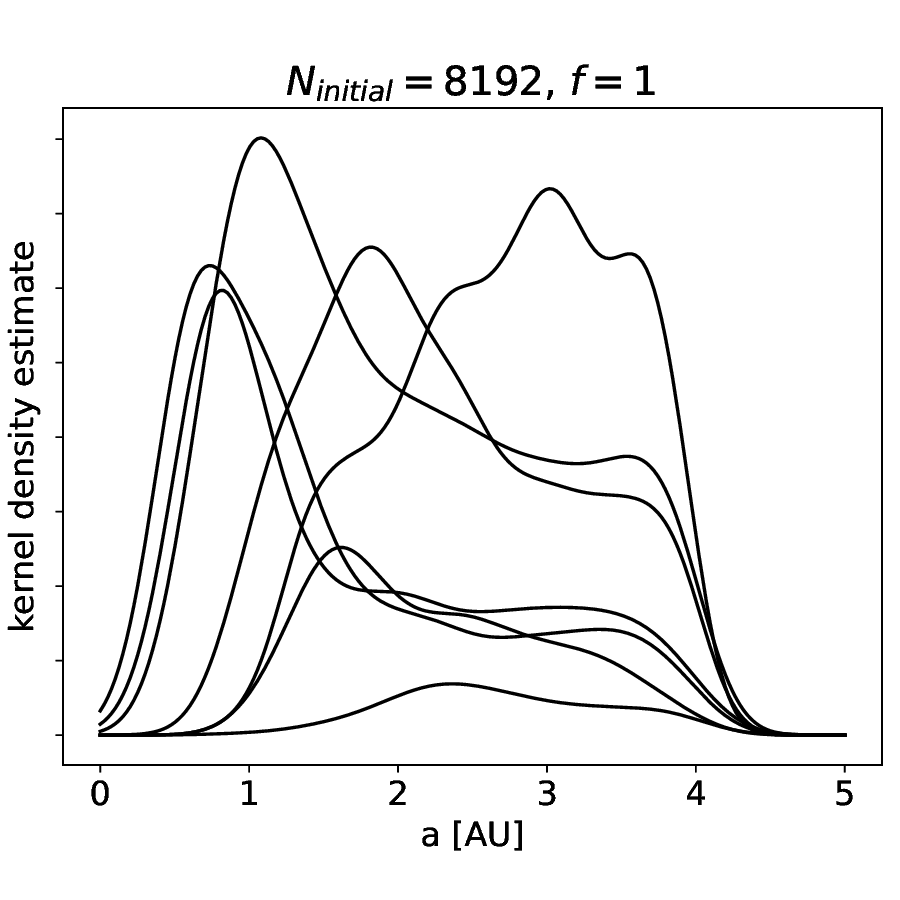}
\includegraphics[width=0.5\textwidth]{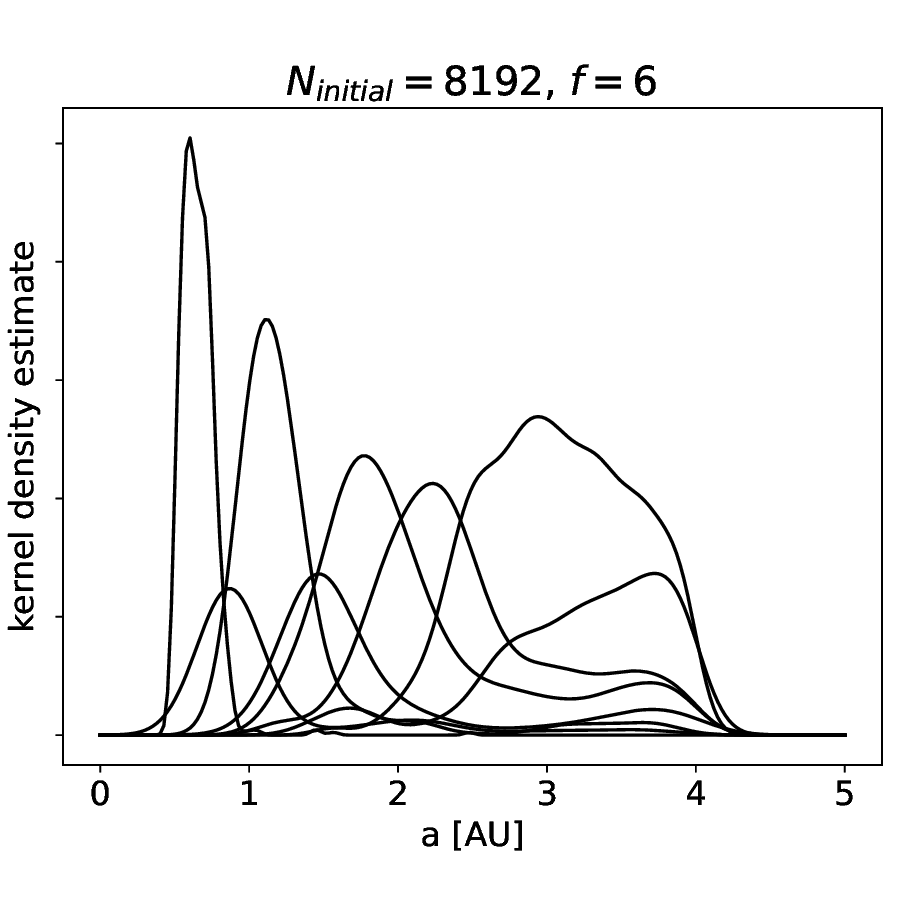}
\caption{Kernel density estimate plot of the initial material location that formed the final planets shown in Figure \ref{fig:ae_f6}. By using an acceleration factor of $f = 6$, the planets get formed mostly by very local material. Without the acceleration factor, the planets composition are more mixed along the entire system. This difference in the composition can have sever consequences in the habitability of the planets and indicate that using an acceleration factor should be avoided.}
\label{fig:composition}
\end{figure}

Modern planetary science goes beyond studying only the masses and sizes of planets. For many exoplanets, the atmospheric composition can be estimated, and even the interior structure of planets is studied. If one is interested in the habitability of planets, then it is important to know the formation history and the interior composition. In $N$-body simulations, the interior composition could be estimated by tracking the initial location of the planetesimals that formed the final planets. Since the initial planetesimal disk can have a radial gradient in composition, this would also impact the composition of the final planets. In Figure \ref{fig:composition}, the initial planetesimal location is shown in the form of a kernel density estimate plot. The figure indicates from what semi-major axis location a planet received its material. One can clearly see a difference between the simulations with and without an acceleration factor. Using $f = 6$, the planets are formed mostly by very local material. Especially, the innermost planets are formed from a very narrow ring of material. The outer planets are a bit more diverse in terms of composition. Without the acceleration factor, all planets are more diverse in terms of the initial material distribution.

This is a very important consequence of the acceleration factor $f$. When one is interested in more than the mass distribution of planetary systems, then this factor should be avoided. Especially because modern hardware and simulation codes are powerful enough to run most simulations without an acceleration factor.

\subsection{Comparing low and high particle resolution simulations}
\label{sec:compareN}

An open question that remains in $N$-body simulations is which particle resolution to use. In \cite{KokuboIda2002}, 10'000 planetesimals are simulated over 400'000 years by using an acceleration factor of six or 10. In \cite{Clement+2020}, 5000 planetesimals are integrated over 100'000 years. In \cite{Raymond+2022}, 1000 planetesimals are integrated over 10 million years. \cite{Brasser2025} used 60'000 planetesimals to simulate the first 5 million years of terrestrial planet formation. And ultimately \cite{Wallace2019} compared a simulation with 4000 planetesimals with a simulation with 1 million planetesimals over a time of 20'000 years, by using an acceleration factor of $f = 6$, to list just a few examples.

\begin{figure}[ht]
\includegraphics[width=\textwidth]{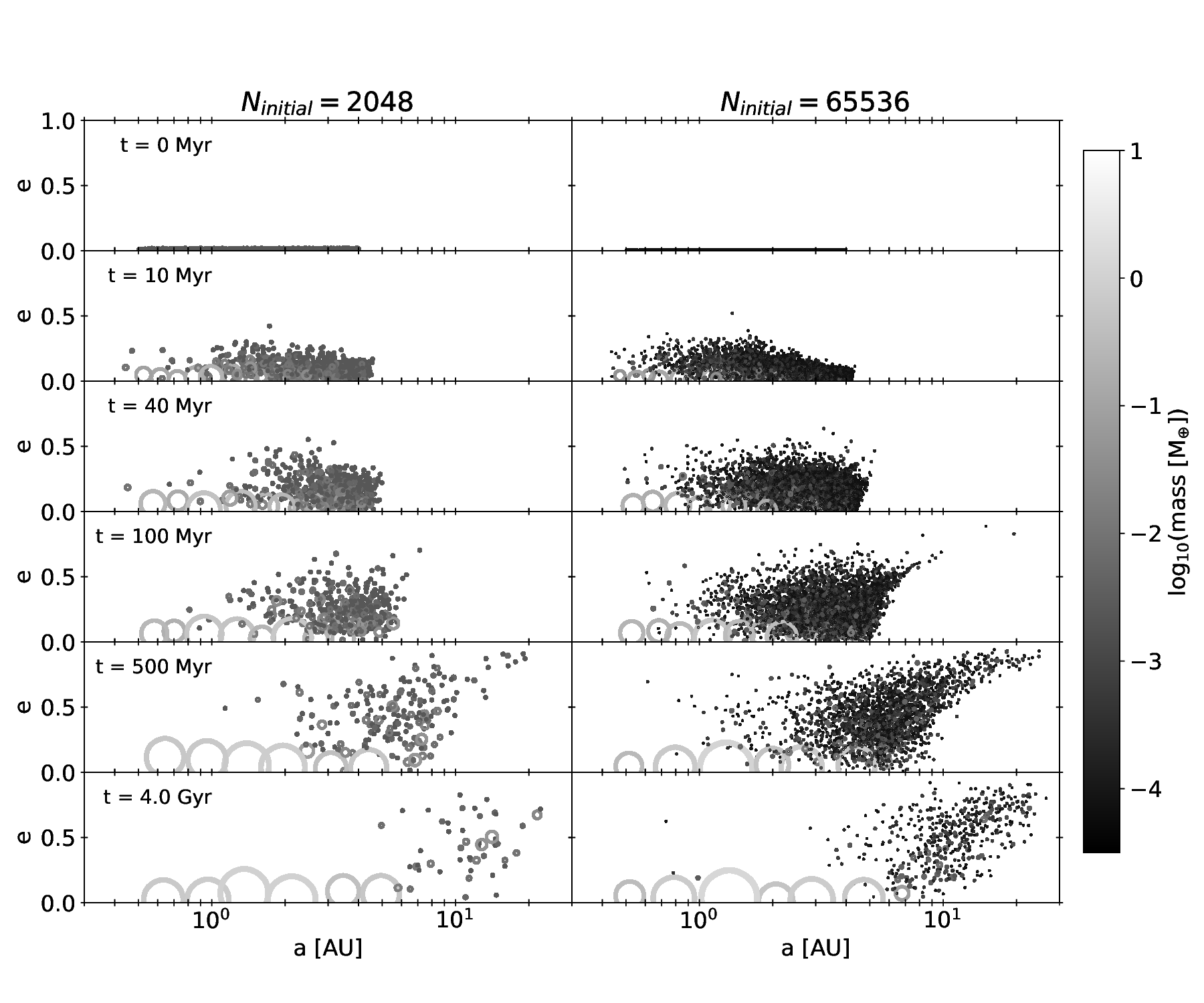}

\caption{Comparing low and high resolution simulations of the formation of terrestrial planets. Both simulations start with $N_{\rm initial}$ planetesimals, distributed in a disk between 0.5 and 4 AU and a total mass of $5 M_{\oplus}$. Shown are the eccentricity $e$ versus the semi-major axis $a$ for different snapshots in time $t$. The color and the size of the symbols represent the mass of the bodies. After 4 Gyr, both simulations have formed six earth like planets in a resonance chain.}
\label{fig:ae1}
\end{figure}

Similarly to \cite{Wallace2019}, we want to study the effect of the particle resolution in terrestrial planet formation simulations. We run simulations with 2048, 4096, 8192, 16384, 32768, and 65536 initial bodies. All simulations run over 4 billion years with GENGA, and in all simulations 6 terrestrial planets are formed with many left-over planetesimals on highly eccentric orbits. In Figure \ref{fig:ae1} are shown snapshots in time of two simulations.
It is very interesting to see that the final configuration of all six simulations with different initial particle numbers show very similar final planets. This is a fact that simulations without perturbing gas giants evolve much more smoothly and are not exposed so much to chaos as shown in \cite{Hoffmann+2017}.

Besides producing the same number of final planets, the bodies appear also in all simulations in a mean motion resonant chain, but the particular resonant period ratio can be different. In Table \ref{tab:P2} are listed the period ratios of all the consequent pairs of bodies. It is important to note that it takes at least 2 Gyr to reach the indicated resonances, and at 4 Gyr, the resonance configurations are not yet fully stable. Especially the weaker configurations still evolve due to interactions with the still present small bodies.

\begin{table}[ht]
\caption{Orbital period ratios of the final planets from different particle resolution simulations. Most of the formed planets are in orbital mean motion resonance configurations near the 2:1 or 5:3 orbital period ratios. It is important to note that the data indicate only the closest strong resonance configuration, the measured period ratio can deviate from these values.}
\label{tab:P2}

\begin{tabular}{p{2cm}p{1cm}p{1cm}p{1cm}p{1cm}p{1cm}}
\hline\noalign{\smallskip}
N \textbackslash Planet pairs & 1-2 & 2-3 & 3-4 & 4-5 & 5-6\\
\noalign{\smallskip}\svhline\noalign{\smallskip}
2048 & 7:4 & 5:3 & 2:1 & 2:1 & 5:3\\
4096 & 7:4 & 5:3 & 7:4 & 2:1 & 8:3\\
8192 & 2:1 & 7:4 & 2:1 & 2:1 & 2:1\\
16384 & 2:1 & 5:3 & 2:1 & 5:3 & 7:4\\
32768 & 2:1 & 5:3 & 2:1 & 5:3 & 7:4\\
65536 & 7:4 & 11:5 & 2:1 & 5:3 & 2:1\\
\noalign{\smallskip}\hline\noalign{\smallskip}
\end{tabular}
\end{table}

\begin{figure}[ht]

\includegraphics[width=0.5\textwidth]{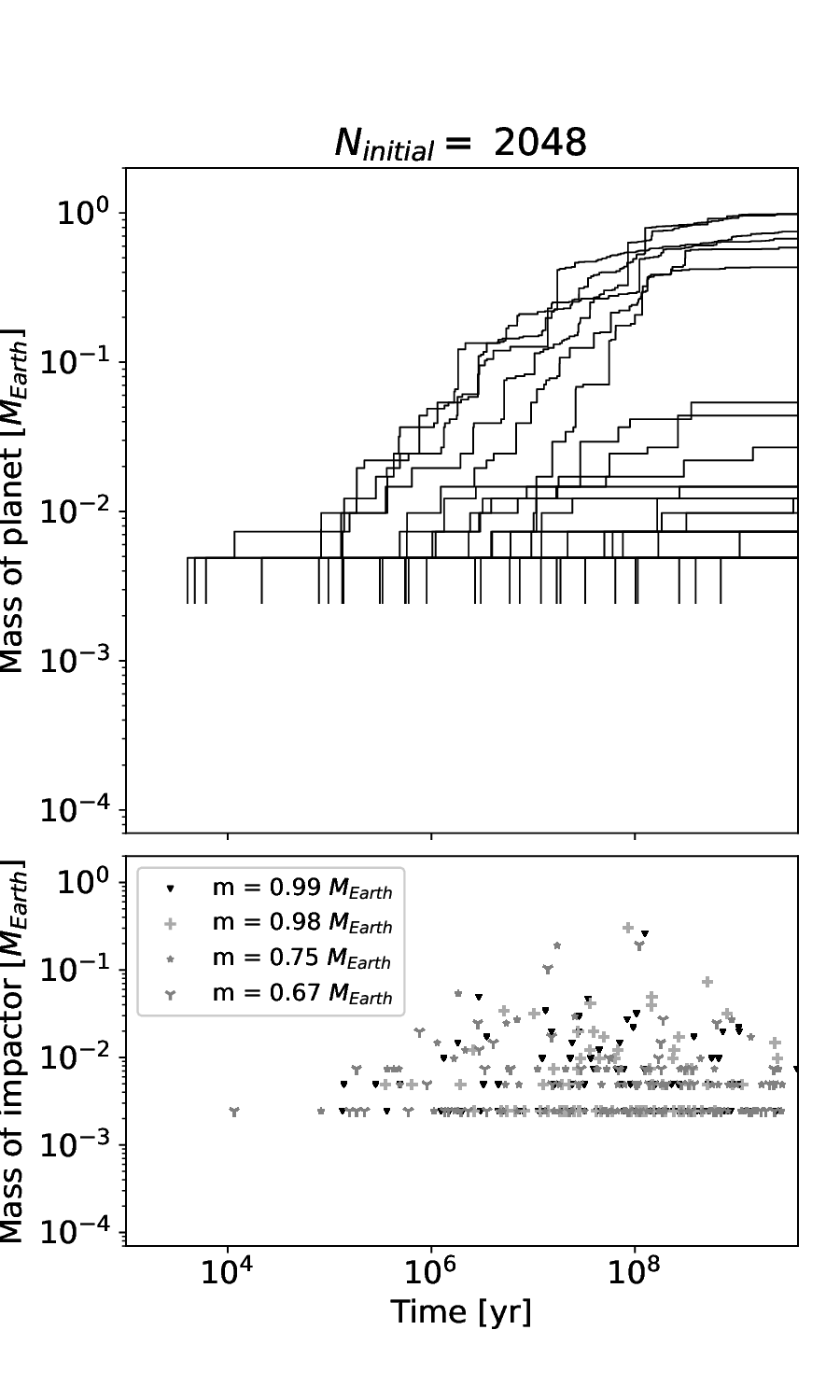}
\includegraphics[width=0.5\textwidth]{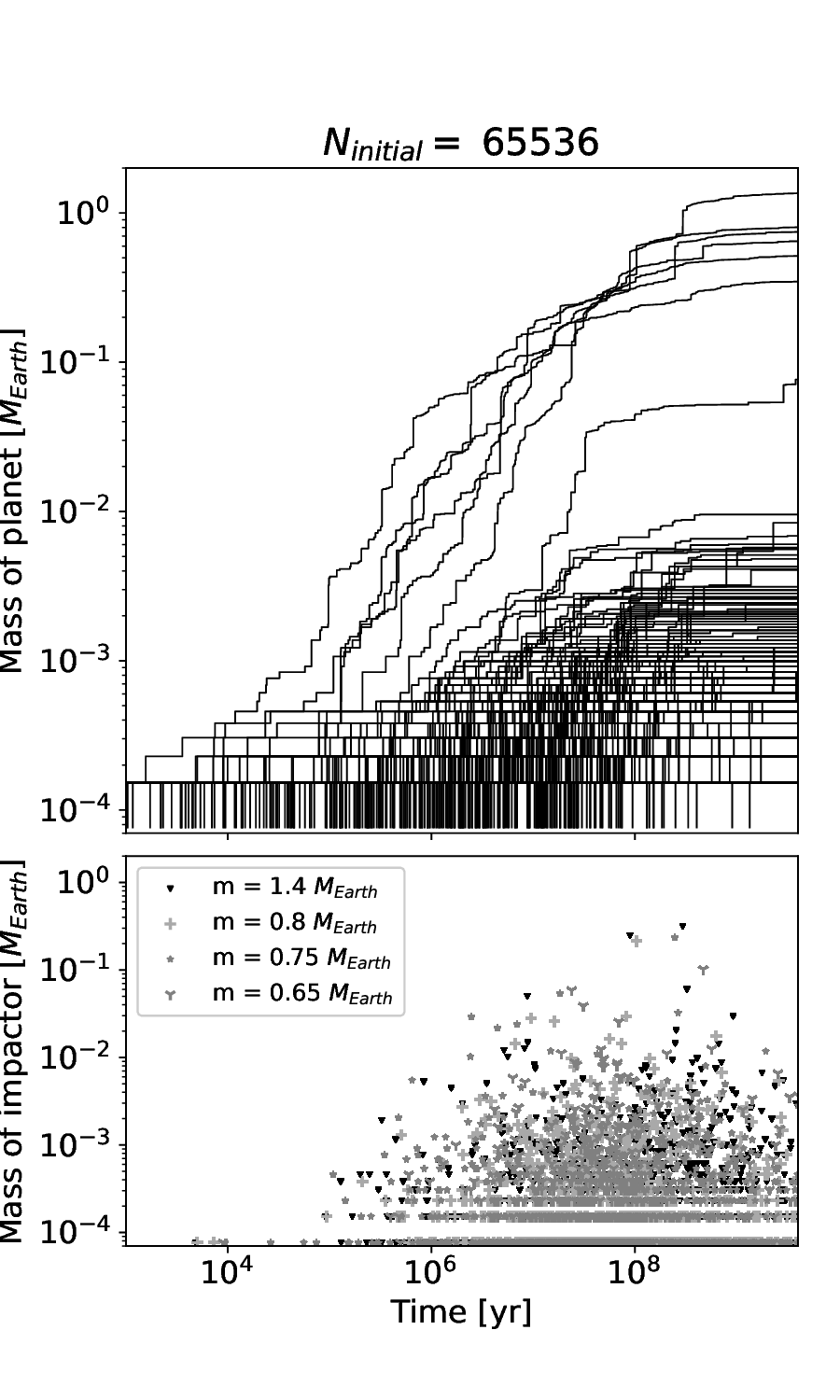}
\caption{Growth tracks of the same simulations as shown in Figure \ref{fig:ae1}. The top panels show how the masses of all remaining bodies grow over time. The bottom plots show all impact events of the four most massive bodies. The growth rate of the planets increases exponentially and culminates each with a giant impact event after $10^7$ - $10^8$ years. After the giant impact event, not much more mass is accreated.}
\label{fig:Coll1}
\end{figure}

Although the final configurations of the planets look very similar across the different particle resolution simulations, there are differences in the formation history of the planets.
This is indicated in Figure \ref{fig:Coll1}. In the top panels, the growth tracks of all remaining bodies after 4 billion years are shown. In the bottom panels, all impact events of the four most massive planets are shown. The data show that in the high-resolution simulation, the growth tracks are in general much more smooth, and most of the impact events come from the smallest bodies around. In both resolution runs, every planet experienced roughly a single giant impact event with an impacter of at least 0.1 Earth masses. The fact that the planets accrete most of the material through smaller planetesimals, in less violent impact events, can change the structure of the planetary interior and also change the composition of the atmosphere. Especially if one extrapolates these results to even higher particle resolutions with even smaller initial planetesimals. 

Another difference between the low- and high-resolution simulations is the formation time of the planetary cores. The low-resolution simulation takes $\approx 2$ Myr to form a 0.1 Earth mass body and $\approx 50$ Myr to form a 0.5 Earth mass body. The high-resolution simulation takes $\approx 4$ and $\approx 90$ Myr for the corresponding masses.

\subsection{Including gas giants}
\begin{figure}[ht]
\includegraphics[width=\textwidth]{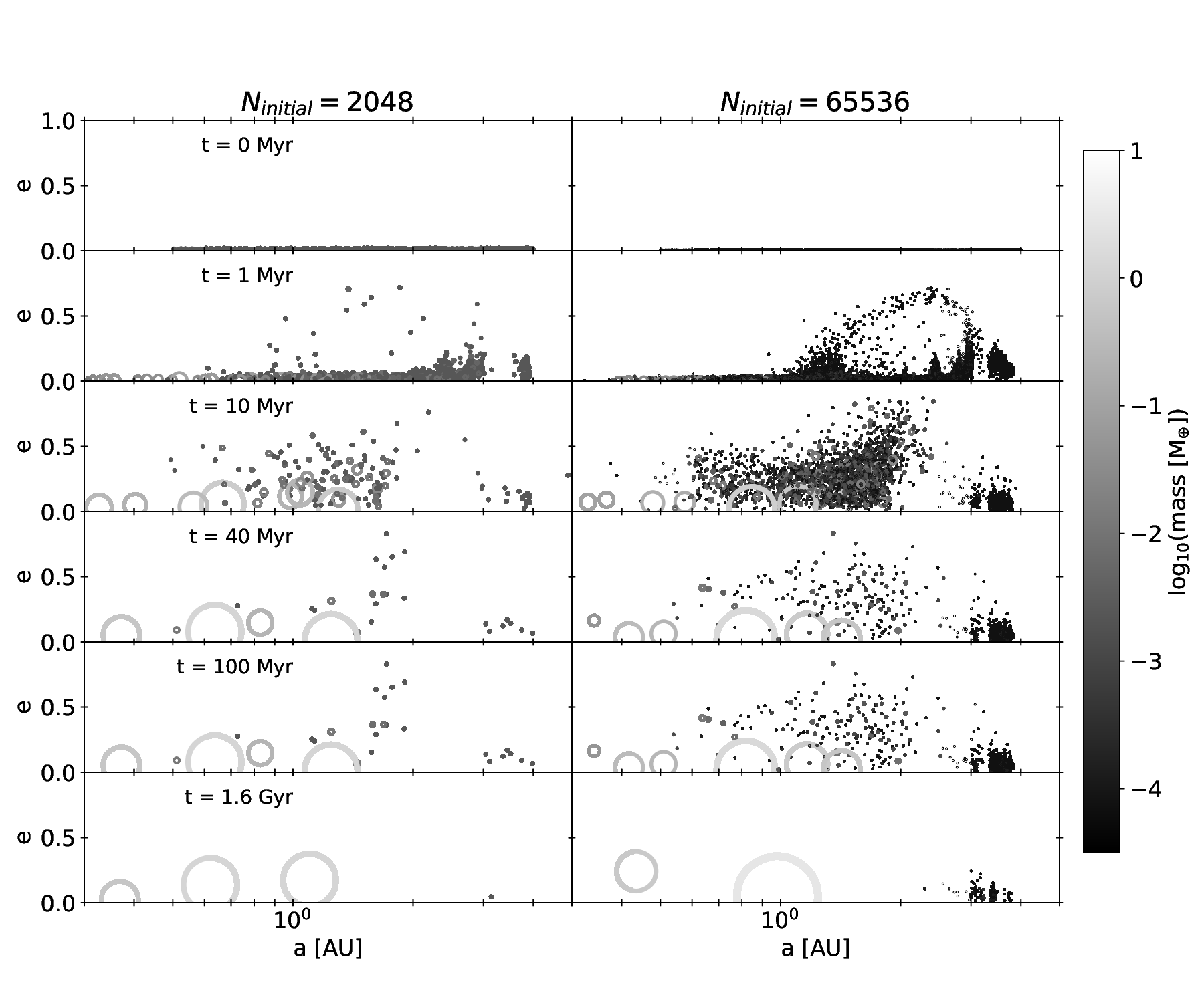}

\caption{Comparing low and high resolution simulations of the formation of terrestrial planets in the presence of Jupiter and Saturn. Both simulations start with $N_{\rm initial}$ planetesimals, distributed in a disk between 0.5 and 4 AU and a total mass of $5 M_{\oplus}$. Shown are the eccentricity $e$ versus the semi-major axis $a$ for different snapshots in time $t$. The color and the size of the symbols represent the mass of the bodies.}
\label{fig:ae2}
\end{figure}

\begin{figure}[ht]
\includegraphics[width=\textwidth]{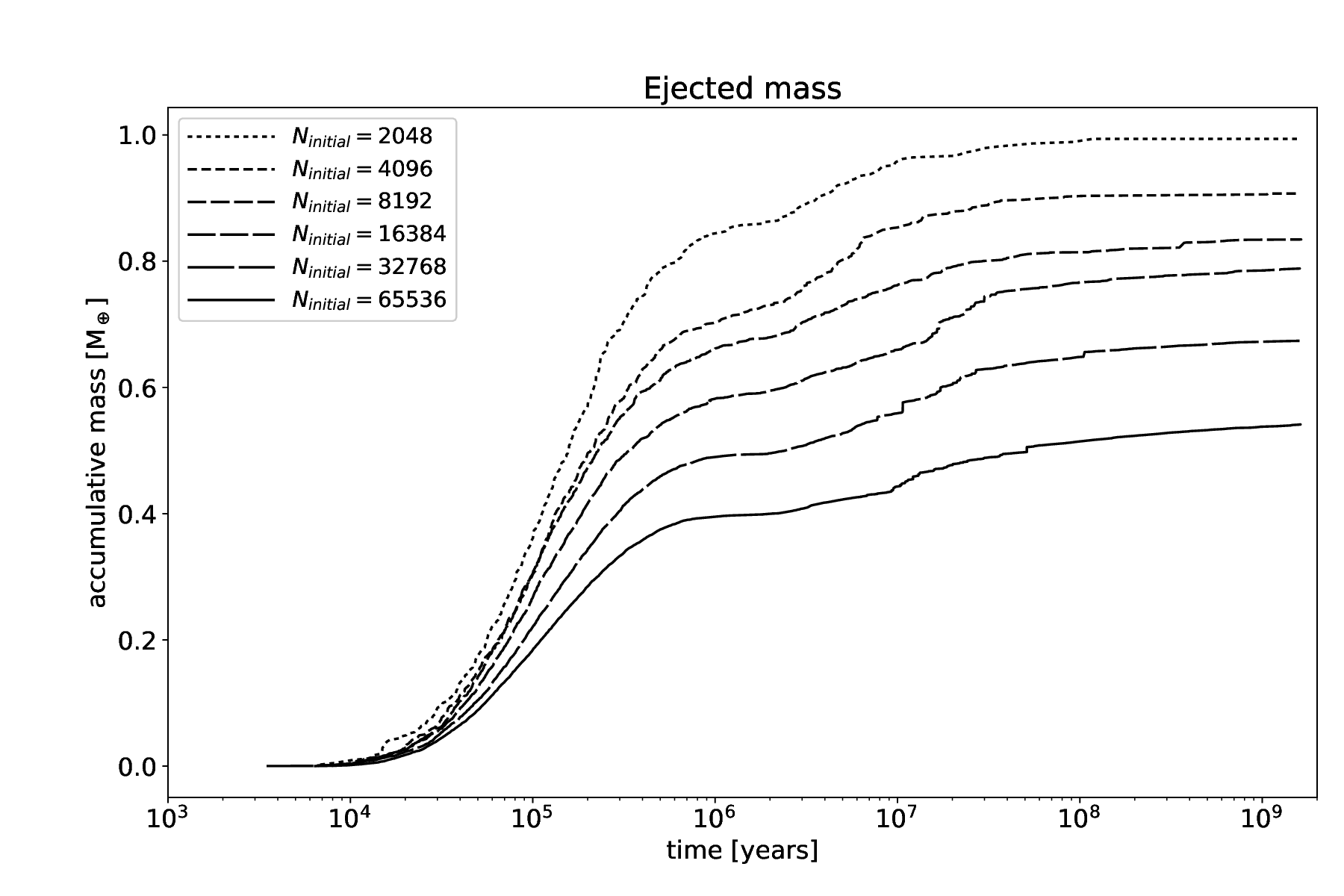}

\caption{Ejected mass of terrestrial planet formation simulations including gas giants and a gas disk with different initial particle resolution. The results show a clear trend towards less ejected masses with higher particle resolutions.}
\label{fig:eject}
\end{figure}

For the case of the Solar System, the situation looks a bit different. 
The presence of gas giants has a major impact on the formation of terrestrial planets in the inner system. The gas giants create strong mean motion resonances, where planetesimals get very eccentric orbits and can therefore collide more easily with other objects. However, also the presence of the gas disk at the beginning of the formation process is important, since the gravitational potential of the gas also creates resonance conditions with the orbiting bodies. As the gas disk vanishes over time due to photoevaporation, the gravitational potential of the gas disk gets reduced and the resonance location of the gas disk moves inward. This so called sweeping resonance line pushes planetesimals to the inner part of the system, where material accumulates and forms terrestrial planets within only a few 10 million years. This process is much more sensitive to chaotic effects than the situation described in the previous section, and the results of individual simulations cannot be compared directly to each other\cite{Hoffmann+2017}. However, it is still interesting to compare the simulations with different particle resolutions. In Figure \ref{fig:ae2} the evolution of a low- and high-resolution simulation is shown. The fact that the simulation with 2048 initial planetesimals formed three terrestrial planets and the simulation with 65536 initial planets formed only two terrestrial planets is a consequence of chaos, and repeating the identical simulations would lead to different configurations. Therefore, studying the number of planets in such simulations would only make sense in a statistical way when many simulations can be run.

However, a different effect is visible only in the high-resolution simulation, that is, the presence of the asteroid belt. Only there, the initial planetesimals are placed fine enough to populate regions between the mean motion resonances and survive over a much longer time. This effect can be important because it allows the delivery of volatile material from the asteroid belt at a later point in time to the already formed planets. 

Similarly to the case without gas giants, there is a difference in the formation time scale between the low- and high-resolution simulations. The low-resolution simulation takes $\approx 1.3$ Myr to form a 0.1 Earth mass body at the inner edge of the system and $\approx 5.5$ Myr to form a 0.5 Earth mass body. The high-resolution simulation takes $\approx 2.8$ and $\approx 9.2$ Myr for the corresponding masses. Focused on the formation of Mars at 1.5 AU, the low-resolution simulation takes 1.6 Myr to form a 0.5 Mars mass body. The high-resolution run takes 3 Myr. Comparing these results with measured Hf-W chronology indicates that the low-simulation matches the data actually better \cite{Woo2021}. However, it must be stated clearly that those results strongly depend on the initial conditions of the planetesimal disk and the used orbital configurations of the gas giants.

Yet another dependency on the resolution of the particles appears in the amount of material ejected from the system as shown in Figure \ref{fig:eject}. The low-resolution simulation with 2048 initial bodies has 1.0 Earth mass of material ejected from the system. The high-resolution run with 65536 initial particles has only 0.5 Earth masses ejected, and the runs in between complete this trend. The shown effect is a combination of the gas effect, acting differently on the different particle sizes, and the fact that in high-resolution runs, less mass is distributed exactly in resonant conditions with the growing planets.

\section{Conclusions}

The use of GPUs in $N$-body simulations has opened new insights into the complex process of terrestrial planet formation. The runtime for large $N$ GPU simulations is much lower than for traditional CPU codes, making it possible to run simulations that resolve finer details of this complex process. The GPU code can also be translated to parallel CPU versions with the aid of cross compilation tools, thereby also making small $N$ simulations more efficient. This combination allows for efficient usage of hardware in situations where the number of particles changes a lot from the start to the end of terrestrial planet formation. Starting with a larger number of lower-mass bodies allows us to begin simulations at an earlier epoch, before the planetary embryos are formed, thus avoiding the need of placing planetary embryos via further model assumptions into the initial conditions. 

The open source code GENGA developed during the NCCR PlanetS offers a powerful $N$-body integrator for planetary systems. It improves the handling of large close encounter groups by improving the hybrid symplectic integration scheme. Originally developed in CUDA for Nvidia GPUs, it offers translators to parallel CPU C++ code using OpenMP and to HIP for running on AMD GPUs. This makes the code very flexible to run. Combined with the implemented non-gravitational effects, it is a state-of-the-art code that is easy to use.

With the presented comparison studies, we show new simulation results on the late stage of terrestrial planet formation. We demonstrate that the acceleration factor $f$ used occasionally in $N$-body simulations to artificially increase the collision time scale should not be used. Even if the final mass distribution looks similar using this acceleration, the formation histories of the planets can be very different. In particular, the initial orbital location of the accreted material can be very different. 

In a particle resolution study, we demonstrate that terrestrial planets can naturally form resonance chains, without the need for orbital migration due to gas effects. The simulations also show that, by using high-resolution simulations, the formation of the planetary cores takes longer as in low-resolution simulations. By increasing the particle resolution from 2048 bodies to 65536 bodies, the formation takes roughly twice as long. In high-resolution runs, most impact events are less violent than in low-resolution runs, which also has an impact on the interior structure as well as on a potential atmosphere composition. This study including gas giants shows a similar trend, namely that the formation of planetary cores takes approximately twice as long in high-resolution runs (65536 bodies)compared to low-resolution runs (2048 bodies). Interestingly, the amount of ejected material from the system in high-resolution runs is only half as large as in low-resolution runs. 

The simulations shown indicate that the ideal particle resolution has not yet been reached and future developments in the $N$-body method focused on even larger $N$ are needed to confirm these new findings. The implementation of the Fast Multipole Method (FMM) into GENGA would allow one to run even larger simulations. Other new developments could include an adaptive time-step method of the hybrid symplectic integration technique allowing to further speed-up simulations involving diverse timescales. Finally, combining the $N$-body method with a more detailed gas disk description and collision modeling could potentially reveal new physical effects critical to the entire planet formation process, especially when the simulation outcome can be linked to observation constraints on the chemical composition and inner structure of planets and asteroids. These results indicate that the development of N-body codes must be pushed forward for larger $N$. Especially the use of realistic fragmentation models in collisions will also increase the necessary $N$. Future integrators should therefore target as a next milestone a million fully interactive particles in a simulation over many billions of years.



\bibliographystyle{spmpsci}

\bibliography{referencesGrimm}

\end{document}